# Determining Research Priorities for Astronomy Using Machine Learning

Brian Thomas[1,2], Harley Thronson[3], Anthony Buonomo[4], Louis Barbier[5]

## Abstract

We summarize the first exploratory investigation into whether Machine Learning (ML) techniques can augment science strategic planning. We find that an approach based on Latent Dirichlet Allocation (LDA) using abstracts drawn from high-impact astronomy journals may provide a leading indicator of future interest in a research topic.

We show two topic metrics that correlate well with the high-priority research areas identified by the 2010 National Academies' Astronomy and Astrophysics Decadal Survey. One metric is based on a sum of the fractional contribution to each topic by all scientific papers ("counts") while the other is the Compound Annual Growth Rate (CAGR) of counts. These same metrics also show the same degree of correlation with the whitepapers submitted to the same Decadal Survey.

Our results suggest that the Decadal Survey may under-emphasize fast growing research. A preliminary version of our work was presented by Thronson et al. (2021).

[1] Corresponding author brian.a.thomas@nasa.gov
[2] NASA Goddard Space Flight Center, 8800 Greenbelt Rd., Greenbelt, MD 20771
[3] 617 Tivoli Passage, Alexandria, VA 22314
[4] Department of Engineering, University of Cambridge, Trumpington St, Cambridge CB2 1PZ
[5] NASA Headquarters, 300 E Street SW, Washington, DC 20546

# 1. INTRODUCTION

One of the most critical planning activities in the sciences is identifying credible priorities for investment. The most high-profile process of scientific prioritization is the National Academies' Decadal Surveys.

A principal challenge faced by this process is the Survey panelists' need to assess a large -- and rapidly growing -- amount of relevant information, specifically many tens of thousands of published research papers. The potential input materials have increased greatly over the years in both variety and quantity, while the basic processes of the Surveys -- and other strategic planning activities -- have changed relatively little. The primary approach for the Surveys over the past half-century (Dressler 2016) remains the same: a central steering committee of a couple dozen members supported by large specialty panels. This leads to the primary motivation of our work: are there ways to substantially improve the current process of identifying the highest-priority science without adding many additional personnel?

We believe that it is time to take advantage of Machine Learning (ML) to augment the daunting task of determining trends and priorities in science from a vast amount of information. Advances in ML over the past decade have been impressive; increasingly powerful ML techniques can comb through a large corpus of unstructured text to reveal insight into their contents. There have recently been examples relevant to the process of science prioritization. For example, Zelnio (2020) reports the successful use of ML to evaluate research literature for promising technologies, and Krenn et al (2019) demonstrate a method used to predict future trends in quantum physics.

# 2. METHODOLOGY AND ANALYSIS

Our goal was to explore whether an AI-based approach would be able to significantly enhance human decision-making with regards to high-impact science research topics. We have created ML models trained on the corpus of scientific research available in advance of the 2010 Decadal Survey. Our process is described more fully in Thronson et al. (2021) and Thomas et al. (2021).

Briefly, Natural Language Processing (NLP) was used to process abstracts and titles drawn from peer-reviewed papers published in the top high-impact journals in astronomy identified per Thomas (2021) during the time period 1998 to 2010. Papers with abstracts of fewer than 100 characters were filtered out of the dataset, leaving ~85,000 abstracts. We utilized NLP to extract scientific terms from the abstracts and use these as features in an algorithm based on Latent Dirichlet Allocation which groups them into research topics.

Measurements of the growth in and relative popularity of the topics may be derived. To determine popularity, we add the fractional contributions that each topic makes to every abstract in the corpus, or "counts". To determine growth rate, we calculate the time series of counts for each topic. These time series may then be analyzed to determine the Compound Annual Growth Rate (CAGR). The "Research Interest" (RI); that is how much overall interest the research community places on a given topic, may be quantified from these measures via

$$RI(t) = (CAGR(t) + 0.1) * counts(t) \qquad (1)$$

where t is the topic.

We next applied these derived topic models to the science frontier panel chapters 1 – 4 for the 2010 Astronomy and Astrophysics Decadal Survey ("2010 corpus"). After cleaning and extracting features as before, we derived counts by topic for each of the paragraphs in the 2010 Survey. Paragraphs where the top three topics contributed less than half of the summed counts total for the were dropped (i.e., these paragraphs did not have good representation by any topic models). A "document content score" (or DCS) for each topic was then derived by taking the remaining paragraphs and summing their counts by topic.

These topic models and associated metrics provide a means to quantify and compare topic content in the literature and the Decadal Survey. We may use this to check for any common relationships and as a check on the validity of these metrics. Figures 1a and 1b show the results of this evaluation.

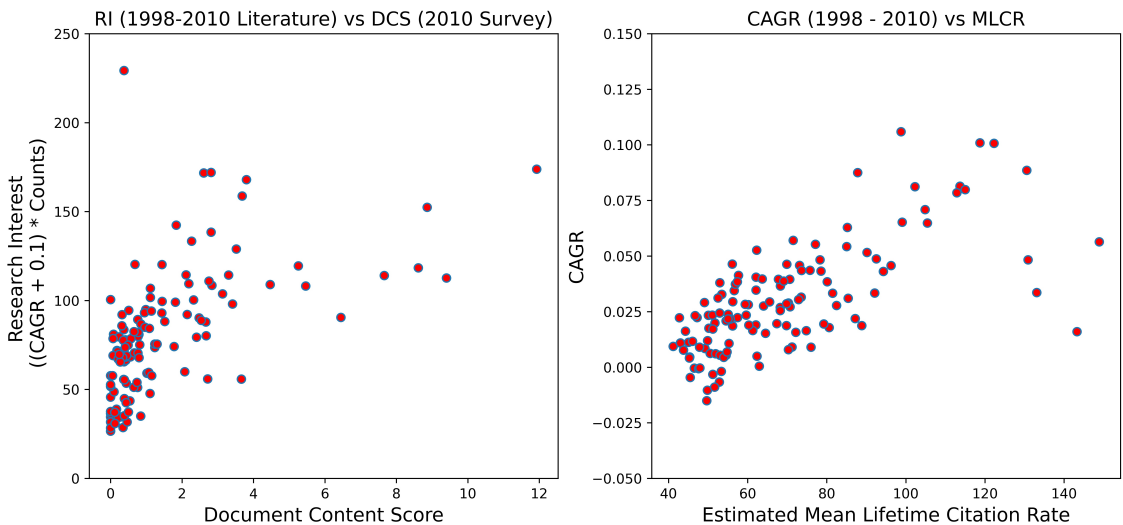

**Figure 1a** (left) The 1998 – 2010 literature RI versus the 2010 Survey DCS by topic (red dots) indicates a significant, but moderate correlation exists. Conversely, only topic CAGR is correlated with the estimated topic Mean Lifetime Citation Rate (MLCR) (**Figure 1b**, right).

Figure 1a compares the RI of published abstracts with the 2010 corpus DCS. We find a highly significant ($P < 0.000001$) correlation of moderate strength ($R \cong 0.6$), which indicates that research which is both growing in interest and/or already has significant research interest is well-represented in the Decadal Survey. A separate analysis of the submitted whitepapers (not shown) also indicates a similar correlation between RI and the content score of the submitted whitepapers.

An essential assumption we have made is that the published body of research accurately reflects the interests and priorities of the community of active astronomers. In order to help ascertain the validity of this assumption we have compared counts, CAGR, and RI for our corpus against the estimated MLCR (Thomas 2021) for these same papers as grouped into each topic. Only CAGR was found to be correlated with the MLCR ($R \cong 0.7$, $P < 0.000001$; see Figure 1b).

## 3. DISCUSSION

Unlike the MLCR, a metric based on citation rates which are a lagging measure of interest in topics of research, these new measures are leading indicators, which makes them attractive for use in planning. There appears to be good, albeit not perfect, correspondence between the frequency of mention of future high-priority research reported in the 2010 Survey and with the content of submitted whitepapers to the RI as determined by the literature of the prior decade. Interestingly, we find only CAGR to be significantly correlated with the estimated MLCR. This result suggests that the Decadal Survey places significant emphasis on <u>established</u> research and may under-emphasize new, growing research topic areas.

We note that in all cases our correlations, although significant, are of only moderate strength and the resultant coefficient of determination ($R^2$), a measure of how much of the variability in one variable can be "explained by" variation in the other, is fairly weak ($R^2 \sim 0.3 - 0.4$). Two reasons may explain why. First, in cases we may be under-sampling the trend in the topic time series, which would lead to some variation in measured CAGR. An alternative issue affects measured DCS: our technique models language present in scientific abstracts, but this may be significantly different from the language present in the 2010 Survey corpus and could sometimes result in lowering the DCS values.

Nevertheless, there is still immediate value in applying this type of analysis to the science prioritization process. The CAGR measure of topics may be exploited to identify probable future impactful research topics and papers, thus creating valuable curated reading. We plan to further understand the variation and uncertainties in Figure 1, which may make it possible to distinguish topic regions in these diagrams and provide additional insight.

ACKNOWLEDGEMENTS

The authors thank Michael Seablom, former Science Mission Directorate Chief Technologist, and Jim Green, former Chief Scientist for their support.